\documentclass[twocolumn,trackchanges]{aastex7}
\newcommand{\edit}[1]{#1}

\usepackage{duckuments}

\usepackage{amsmath}

\defcitealias{Rui:2024:cv-mergers}{Paper I}
\defcitealias{Kippenhahn:1980:Thermohaline}{KRT80}
\defcitealias{Brown:2013:Thermohaline}{BGS13}

\begin{document}

\title{Forging neon-distilling white dwarfs in the stellar engulfments of helium white dwarfs}

\author[orcid=0000-0002-1884-3992]{Nicholas Z. Rui}
\altaffiliation{NASA Hubble Fellow}
\affiliation{Department of Astrophysical Sciences, Princeton University, 4 Ivy Lane, Princeton, NJ 08544, USA}
\affiliation{Center for Interdisciplinary Exploration and Research in Astrophysics (CIERA), Northwestern University, 1800 Sherman Ave., Evanston, IL 60201, USA}
\email[show]{nrui@princeton.edu}

\author[orcid=0000-0002-4544-0750]{Jim Fuller}
\affiliation{TAPIR, California Institute of Technology, Pasadena, CA 91125, USA}
\email{jfuller@caltech.edu}


\begin{abstract}
Once carbon--oxygen white dwarfs cool sufficiently, they crystallize from the inside out. If the white dwarf is rich enough in ${}^{22}\mathrm{Ne}$, these crystallized solids are buoyant and rapidly rise, efficiently liberating potential energy which may halt the cooling of the white dwarf or power magnetic phenomena.
Although this ${}^{22}\mathrm{Ne}$ distillation process may explain the cooling anomaly in Q-branch white dwarfs and anomalous emission lines in DAHe white dwarfs, its operation demands unusually high ${}^{22}\mathrm{Ne}$ abundances not generically predicted by isolated stellar evolution.
We show that the engulfments of helium white dwarfs by both main-sequence and red giant stars can result in carbon--oxygen white dwarfs with ${}^{22}\mathrm{Ne}$ abundances high enough to distill ${}^{22}\mathrm{Ne}$.
This enhancement occurs because carbon dredged up following an especially energetic and off-center helium flash can be processed into ${}^{22}\mathrm{Ne}$ by subsequent hydrogen shell burning and helium shell burning.
${}^{22}\mathrm{Ne}$-distilling white dwarfs from these merger channels are predicted to be somewhat more massive than typical white dwarfs (up to $\simeq0.7M_\odot$) and may have anomalous rotation rates, consistent with DAHe white dwarfs. These binary formation channels for ${}^{22}\mathrm{Ne}$-rich white dwarfs reveal new connections between binary interactions and white dwarf cooling phenomena.
\end{abstract}


\section{Introduction}

\begin{figure*}
    \centering
    \includegraphics[width=\linewidth]{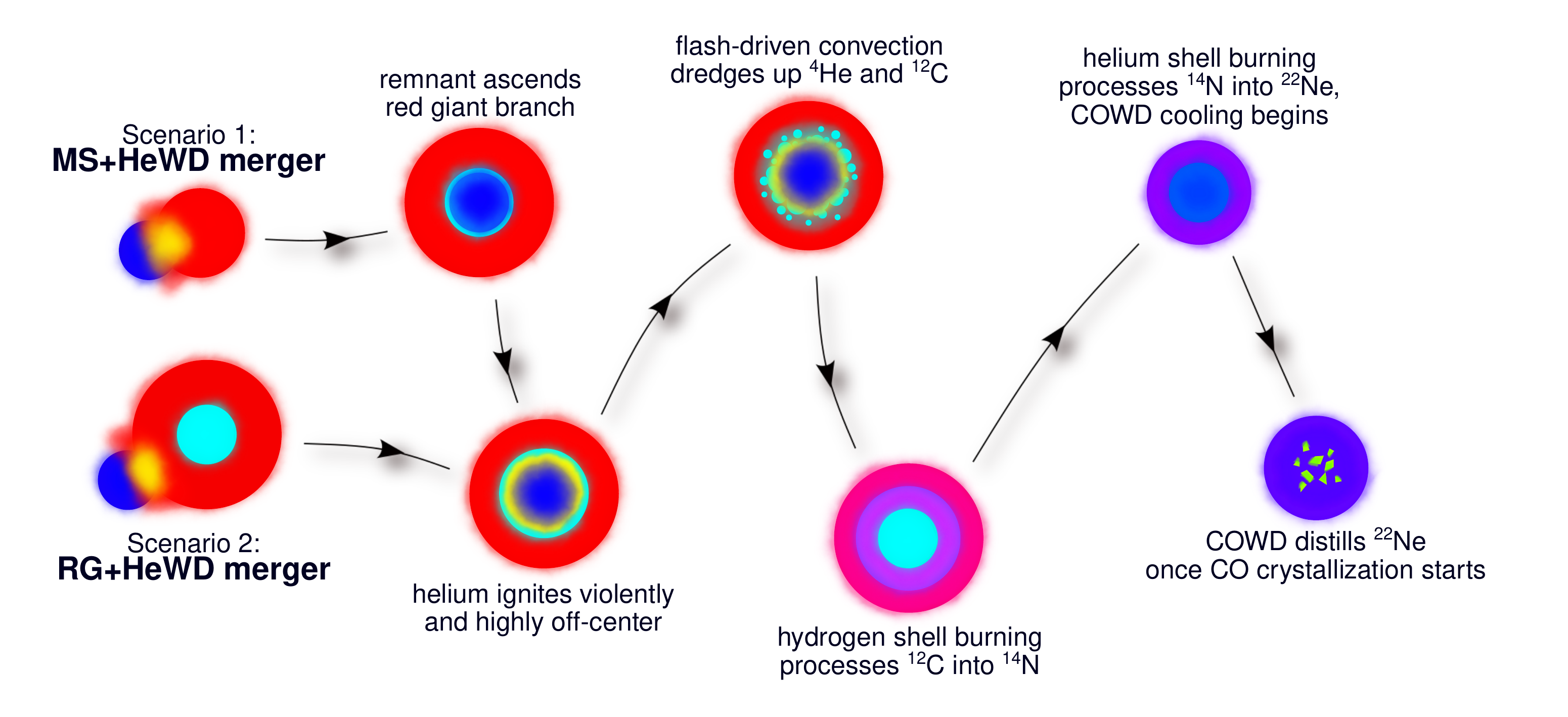}
    \caption{A cartoon outlining the MS+HeWD and RG+HeWD merger channels for forming COWDs which are capable of ${}^{22}\mathrm{Ne}$ distillation.
    In both cases, a mixing event following a highly off-center helium flash dredges up a substantial amount of ${}^{12}\mathrm{C}$, which can later be processed into ${}^{22}\mathrm{Ne}$.}
    \label{fig:sbr_cartoon}
\end{figure*}

White dwarfs (WDs) are the most ubiquitous species of compact object, typically the final stages of stars with masses $M\lesssim8M_\odot$.
Since they no longer host substantial nuclear burning, WD evolution primarily consists of cooling through radiating their remaining reservoir of thermal energy into space.

When their inner layers cool sufficiently, WDs undergo phase transitions (``crystallization'') which stall their cooling through the release of both latent heat and gravitational potential energy \citep{vanHorn:1968:WDcrystallization}.
The canonical WD has a mass $\simeq0.6M_\odot$ and a carbon--oxygen (CO) composition and crystallizes at roughly $T_{\mathrm{eff}}\simeq6\,\mathrm{kK}$ \edit{\citep{Bedard:2020:STELUM,Bauer:2023:COPhaseSeparation}}, with more massive WDs crystallizing at higher effective temperatures.
At the onset of crystallization, the CO plasma mixture at the center of the star separates into an oxygen-rich solid phase and a carbon-rich liquid phase.
For typical WD compositions, the heavier solid phase accumulates at the center of the WD while the liquid phase buoyantly rises into the overlying fluid.
The resulting convective or thermohaline mixing has been proposed as a mechanism for creating magnetic WDs, either by driving a magnetic dynamo \citep{Isern:2017:CrystallizationDynamo,Ginzburg:2022:CrystallizationDynamo,Fuentes:2024:ShortIntenseDynamo} or at least surfacing pre-existing buried magnetic fields \citep[e.g.,][]{CastroTapia:2024:BDiffusion,CastroTapia:2025:BDiffusion,Blatman:2024:BDiffusion,Blatman:2024:UMBDiffusion,Blatman:2025:UMWDseismology}.

However, if the COWD is sufficiently enhanced in the neutron-rich species ${}^{22}\mathrm{Ne}$, the solid phase (now poor in ${}^{22}\mathrm{Ne}$) has a lower density than the original plasma and, as a result, floats \citep{Segretain:1996:NeonDistillation,Blouin:2021:QBranch}.
Crystallization then causes the rapid, buoyant rise of oxygen-rich crystals accompanied by the concentration of ${}^{22}\mathrm{Ne}$-rich material at the center of the WD (so-called ``${}^{22}\mathrm{Ne}$ distillation'').
Since ${}^{22}\mathrm{Ne}$ distillation concentrates heavy, neutron-rich atoms in the COWD's center more efficiently than standard CO crystallization, it is capable of stalling the WD's cooling for much longer \citep[$\gtrsim7\,\mathrm{Gyr}$;][]{Bedard:2024:Buoyant}.
As a result, ${}^{22}\mathrm{Ne}$ distillation has been invoked to explain the overdensity of WDs in the ultramassive Q branch \citep[with masses $\gtrsim1.0M_\odot$;][]{Cheng:2019:QBranch,Bauer:2020:NeonClustering,Blouin:2021:QBranch,Bedard:2024:Buoyant} as well as the unusual Balmer emission lines observed in DAHe-type WDs \citep[with masses $\simeq0.8M_\odot$;][]{Lanza:2024:DAHes}.

Prompt ${}^{22}\mathrm{Ne}$ distillation only occurs for high initial ${}^{22}\mathrm{Ne}$ abundances \citep[e.g., $X({}^{22}\mathrm{Ne})\gtrsim3\%$ for $X_{\mathrm{O}}\simeq70\%$;][]{Blouin:2021:QBranch}.
This threshold is much higher than the ${}^{22}\mathrm{Ne}$ abundances predicted by canonical stellar evolution \citep[$X({}^{22}\mathrm{Ne})\simeq1.5$--$2\%$;][]{Salaris:2022:BASTIisochrones}.
It is also not met by the vast majority of WDs in the solar neighborhood, based on detailed modeling of their CNO abundances \citep{Barrientos:2025:NeHabitableZone}.
${}^{22}\mathrm{Ne}$ is produced primarily from ${}^{14}\mathrm{N}$ experiencing two successive $\alpha$ captures separated by a $\beta$ decay.
In turn, $^{14}\mathrm{N}$ is generated in CNO-cycle hydrogen-burning regions from other metals participating in the CNO cycle (e.g., ${}^{12}\mathrm{C}$).
The required ${}^{22}\mathrm{Ne}$ mass fraction may be inherited from an abnormally $\alpha$-rich progenitor \citep{Salaris:2024:NeonOpenCluster} or, alternatively, produced during a stellar merger.
In the latter case, \citet[][]{Shen:2023:Subgiant} have shown that mergers between COWDs and subgiants may produce enough ${}^{22}\mathrm{Ne}$ (together with ${}^{26}\mathrm{Mg}$, another neutron-rich species) to undergo distillation.
This raises the intriguing prospect of distillation-induced phenomenology in WDs as a probe of binary interactions in earlier phases of stellar evolution.

In this work, we investigate the COWD remnants of stellar mergers involving helium WDs (HeWDs), the stripped helium-rich cores of red giant stars.
Because HeWDs cool in the same manner as COWDs, stellar mergers involving them imply the ingestion of a substantial amount of dense, very low-entropy helium.
Such mergers often produce red-giant-like stellar merger remnants which behave differently than single stars.
\edit{In particular, these merger remnants often degenerately ignite helium substantially more energetically and farther off-center than a single star would.
In many cases, these atypical helium flashes} induce the dredge up of a substantial amount of helium, carbon, and other species \edit{from the core} into the convective envelope (\citealt{Zhang:2013:EarlyRJ}, \citealt{Zhang:2017:evolution}, and \citealt{Rui:2024:cv-mergers}, hereafter \citetalias{Rui:2024:cv-mergers}).
Notably, \citetalias{Rui:2024:cv-mergers} finds that the envelopes of remnants of mergers between main-sequence (MS) stars and HeWDs become rich in ${}^{22}\mathrm{Ne}$ (see also Figure 13 of \citealt{Zhang:2013:EarlyRJ}, in the case of red giants; RGs).

In this work, we show that the remnants of MS+HeWD and RG+HeWD mergers can evolve into COWDs which are ${}^{22}\mathrm{Ne}$-rich enough to undergo ${}^{22}\mathrm{Ne}$ distillation immediately upon crystallizing.
The process by which MS+HeWD and RG+HeWD merger remnants eventually evolve into ${}^{22}\mathrm{Ne}$-rich COWDs is schematically described in Figure \ref{fig:sbr_cartoon}.

\section{Models}

Capturing the precise physics of stellar mergers generally requires computationally intensive hydrodynamical simulations \citep[][]{Lombardi:2002:MMAS,Schneider:2016:MagneticMergers,Wu:2020:StellarMergers}.
Fortunately, as this work concerns the long-term evolution of the merger remnant long after it has settled into hydrostatic equilibrium, we can inexpensively ``stellar engineer'' plausible merger remnant models under plausible assumptions.

We create and evolve our merger remnant using version r24.08.1 of the Modules for Experiments in Stellar Astrophysics (\texttt{MESA}) code \citep{Paxton:2011:MESA,Paxton:2013:MESA,Paxton:2015:MESA,Paxton:2018:MESA,Paxton:2019:MESA,Jermyn:2023:MESA} in tandem with custom Python routines.
We use \texttt{MESA}'s \texttt{cno\_extras\_o18\_to\_mg26.net} nuclear reaction network and initialize our models with solar metallicity using the metal fractions of \citet{Grevesse:1998:SolarComposition}, although we manually remove some trace elements during certain steps of the stellar engineering process in order to avoid unphysical burning events.
We also selectively disable compositions due to burning and mixing during some dynamical-timescale phases which are artificially extended in our \texttt{MESA} modeling.
Wind during the asymptotic giant branch (AGB) is modeled as a Bl\"ocker wind with $\eta_B=0.1$ \citep{Bloecker:1995:AGBwind}.


In this work, we model the remnants of MS+HeWD mergers and RG+HeWD mergers.
As our stellar engineering methods involve many steps in order to generate both physically realistic and numerically convergent models, we describe the general outline of our procedures and make the files required to reproduce our results publicly available\footnote{\texttt{MESA} files used to create and evolve our stellar models can be found at the following \edit{URL: \href{https://zenodo.org/records/17980321}{https://zenodo.org/records/17980321}}.
The code used to automatically generate our \texttt{MESA} setups used is stored in the following repository: \edit{\href{https://github.com/NicholasRui/qol}{https://github.com/NicholasRui/qol}}.}.
In order to focus on structural and chemical properties of merger remnants, we do not include the effects of rotation, which may be anomalous in merger remnants \citep[e.g.,][]{Tayar:2015:RapidRotation}.
In addition to our merger remnant models, we also evolve a $1.2M_\odot$ star from the zero-age main sequence to the COWD phase with identical physics in order to compare our results to an outcome of isolated stellar evolution.

In this work, we do not explicitly include winds on the red giant branch (RGB), instead implicitly including this mass loss into $M_{\mathrm{env}}$.
This omission is most relevant for the MS+HeWD case, where the merger remnant spends a substantial amount of time near the tip of its RGB (see \citealt{Zhang:2017:evolution} and \citetalias{Rui:2024:cv-mergers}).
When including a Reimers wind \citep{Reimers:1975:RGBwind} with scaling factor $\eta_R=0.5$, \citet{Zhang:2017:evolution} find that MS+HeWD merger remnants lose the majority of their envelopes, only reaching the core-helium-burning phase if $M_{\mathrm{MS}}\gtrsim0.6M_\odot$.
However, using seismic mass measurements of RGs, \citet{Li:2025:RGBMassLoss} find that RGB mass loss weakens at higher metallicities, consistent with $\eta_R=0.1$ at solar metallicity (their Figure 12).
As a test, we re-evolve the HeWD+MS merger remnant model with fiducial parameters $M_{\mathrm{HeWD}}=0.4M_\odot$ and $M_{\mathrm{env}}=0.4M_\odot$ with both $\eta_R=0.5$ and $\eta_R=0.1$.
Consistent with expectations, in the former case, the star loses most of its envelope to winds ($\approx0.32M_\odot$) and fizzles out into a $\approx0.47M_\odot$ HeWD.
In the latter case, the star only loses $\lesssim0.1M_\odot$, comfortably reaching the core-helium-burning phase.
Moreover, since MS+HeWD merger remnants typically reach brighter luminosities than normal RGs before igniting helium \citepalias{Rui:2024:cv-mergers}, their mass loss could be overestimated by prescriptions calibrated to largely isolated stars.
We conclude that it is at least plausible if not likely for MS+HeWD merger remnants to retain a large fraction of their envelopes during the RGB.

\subsection{MS+HeWD merger remnants} \label{sect:making_mshewd}

\begin{figure*}
    \centering
    \includegraphics[width=\textwidth]{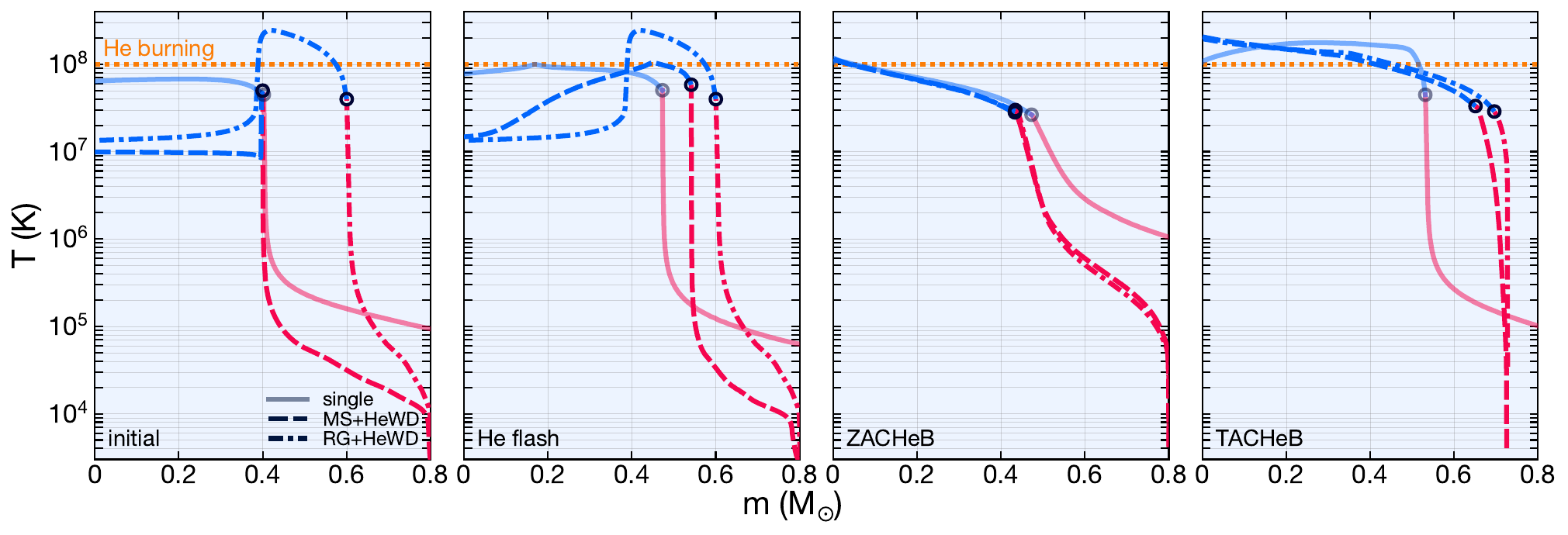}
    \caption{Temperature profiles for two representative $M=0.80M_\odot$ merger remnants: a MS+HeWD merger remnant with $M_{\mathrm{HeWD}}=0.40M_\odot$ and $M_{\mathrm{env}}=0.40M_\odot$ (dashed) and a RG+HeWD (somewhat non-conservative) merger remnant with $M_{\mathrm{HeWD}}=0.40M_\odot$, $M_{\mathrm{RGcore}}=0.20M_\odot$, and $M_{\mathrm{env}}=0.20M_\odot$ (dash-dotted).
    Profiles are shown soon after the merger (first column from left), immediately before degenerate helium ignition (second), at the onset of core helium burning (third), and once core helium has been exhausted (fourth).
    The \textit{red} and \textit{blue} portions of the curves denote the hydrogen-rich envelope and helium core, with the black circle denoting the boundary between them (defined by $X=10^{-3}$).
    The dotted orange line indicates the approximate temperature for helium burning ($T\approx10^8\,\mathrm{K}$).
    The inner mass coordinates of a $1.2M_\odot$ isolated RG model (starting from $M_{\mathrm{core}}=0.40M_\odot$) are shown for comparison (translucent solid).
    }
    \label{fig:rg_profiles}
\end{figure*}

Observationally, there is a dearth of currently accreting HeWDs, despite predictions from population synthesis that they should be common \citep{Zorotovic:2011:WDMassProblem,Zorotovic:2020:WDMassProblem,Schreiber:2016:CAML}.
This fact has been used to imply that mass transfer from low-mass MS stars onto HeWDs is typically unstable, and that such systems quickly merge after the onset of accretion.
Although still uncertain, the mass transfer destabilization mechanism may be consequential angular momentum loss caused by frictional interaction with a slowly expanding nova shell \edit{\citep{Shen:2015:SlowNovaMergers,Nelemans:2016:NovaInteraction,Schreiber:2016:CAML}}.

In the resulting merger, the MS star disrupts around and accretes onto the HeWD, whose structure is unlikely to be affected due to its much higher density and long internal thermal time \citepalias[$\sim10$s--$100\,\mathrm{Myr}$;][]{Rui:2024:cv-mergers}.
As long as the accretion efficiency is not very low (so that $\gtrsim10^{-3}M_\odot$ of material successfully accretes), the hydrogen-rich material undergoes hydrogen shell burning and inflates into an extended envelope.
Although such merger remnants are similar to normal RGs, they possess unusually low-entropy cores due to the extended period of cooling experienced by the progenitor HeWDs.
This often produces significant temperature inversions in the core, in contrast to the minor neutrino-cooling-induced temperature inversions in the cores of normal RGs.

To create a model of a MS+HeWD merger remnant, we first generate a HeWD model with mass $M_{\mathrm{HeWD}}$ by removing the envelope of an RG model and evolving the resulting model down a cooling track until it reaches $T_{\mathrm{eff}}\approx10\,\mathrm{kK}$.
We then separately initialize a main-sequence model with the desired mass $M_{\mathrm{env}}$ of the accreted envelope and relax its inner boundary condition so that the radius and enclosed mass of the innermost cell matches that of the HeWD model.
We then manually ``stitch'' the two models together and evolve the resulting model with hydrodynamics enabled in order to ``ring down'' the resulting model into hydrostatic equilibrium.
This phase is kept short ($10\,\mathrm{kyr}$) in order to roughly preserve the entropy profile of the core, which does not thermalize much over this timescale.
The merger remnant is subsequently evolved in hydrodynamical mode in order to accommodate radial pulsations which occur at the tip of the RGB.

Under this procedure, we do not model any merger-induced mixing between the core and the envelope \citep[e.g., due to shear instabilities;][]{MacDonald:1983:ShearMixing}.
Moreover, for the subsequent analysis, we implicitly assume an order-unity fraction of the MS star's material successfully accretes, so that the total mass of the merger remnant is sufficiently large to initiate helium burning ($M_{\mathrm{rem}}\gtrsim0.5M_\odot$).
Although \citet{Metzger:2021:cv-deaths} find that the highly super-Eddington disk produced by the disrupted secondary can experience outflows which expel most of the mass, they also note that effects such as gravitational instabilities may allow more mass to accrete.
Nevertheless, this remains a significant uncertainty.

This procedure is qualitatively similar to that of \citetalias{Rui:2024:cv-mergers}, which stitches a HeWD model onto a convective envelope taken from a RG model (rather than creating it by relaxing the initial condition of a MS model).
Because a hot RG core has a somewhat larger core than a cooled HeWD, we find that this improved procedure creates an envelope which is a better fit to the HeWD, and thus numerically accommodates higher-mass HeWDs.
The left column of Figure \ref{fig:rg_profiles} shows a MS+HeWD merger remnant constructed using fiducial parameters $M_{\mathrm{HeWD}}=0.40M_\odot$ and $M_{\mathrm{env}}=0.40M_\odot$.

In addition to the fiducial model, we also present a model with $M_{\mathrm{HeWD}}=0.45M_\odot$ and $M_{\mathrm{env}}=0.4M_\odot$ (i.e., near maximal-mass $M_{\mathrm{HeWD}}$) and a model with $M_{\mathrm{HeWD}}=0.40M_\odot$ and $M_{\mathrm{env}}=0.60M_\odot$ (i.e., a more massive envelope).
These models are summarized in Table \ref{tab:mshewd_table}. 

\begin{table}
    \centering
    \begin{tabular}{l l | l l}
        \hline
        $M_{\mathrm{HeWD}}$ & $M_{\mathrm{env}}$ & $M_{\mathrm{COWD}}$ & $X({}^{22}\mathrm{Ne})_{\mathrm{tot}}$ \\
        \hline
        $0.40$ & $0.40$ & $0.66$ & $2.95\%$ \\
        $0.45$ & $0.40$ & $0.70$ & $3.53\%$ \\
        $0.40$ & $0.60$ & $0.60$ & $2.47\%$ \\
        \hline
    \end{tabular}
    \caption{Initial conditions and final mass and ${}^{22}\mathrm{Ne}$ mass fraction of our MS+HeWD merger remnant models.
    All masses are given in $M_\odot$.}
    \label{tab:mshewd_table}
\end{table}

\subsection{RG+HeWD merger remnants} \label{sect:making_rghewd}

RG-HeWD mergers have been invoked as a promising channel for the formation of early-type R and J class carbon stars \citep{Izzard:2007:EarlyR,Zhang:2013:EarlyRJ,Zhang:2020:PopSynth}, as a possible progenitor for the transient CK Vulpeculae \citep{Tylenda:2024:CKVul}, and as a possible explanation for the unusual 8 UMi planetary system \citep{Hon:2023:8UMi}.
They are natural outcomes of close binary evolution in which a first phase of mass transfer (which creates the HeWD) is followed by a common-envelope merger during a second phase of mass transfer.

Unlike in the MS+HeWD case in which one object is much denser than the other, in RG+HeWD mergers the degenerate helium core of the RG is comparable in density to the HeWD.
Construction of the merger remnant therefore requires an assumption about how these two components combine to create the merger remnant's core.
Testing multiple different scenarios, \citet{Zhang:2013:EarlyRJ} find that they can only \edit{reproduce} the properties of early-R carbon stars \edit{with a high-mass HeWD ($M_{\mathrm{HeWD}}\simeq0.4M_\odot$) and low-mass RG core.
To reproduce early-R stars, they further must assume that the merger remnant's core consists of the mostly intact HeWD surrounded by marginally degenerate helium from the RG core.}
This scenario is most conducive to an especially energetic, off-center helium flash which causes the core dredge-up event required for significant ${}^{22}\mathrm{Ne}$ production.

\edit{\citet{Zhang:2013:EarlyRJ} imagine this core structure to result from the subduction of the low-entropy HeWD beneath the high-entropy RG core, consistent with heuristic expectation that adiabatic stellar mergers should produce approximately entropy-sorted remnants (e.g., \citealt{Lombardi:2002:MMAS}, although ignoring shock heating, e.g., \citealt{Gaburov:2008:MMAMS}).
However, mergers between HeWDs and RG cores instead likely resemble double WD mergers, for which hydrodynamical simulations consistently show the disruption and subsequent accretion of the lower-mass WD \citep{LorenAguilar:2009:WDMergers,vanKerkwijk:2010:WDMergers,Raskin:2012:WDMergers,Pakmor:2012:WDMergers,Dan:2012:WDMergers,Dan:2014:WDMergers,Pakmor:2024:HeWDMerger}.
A merger between a HeWD and RG core therefore more likely proceeds via the disruption of the RG core followed by its accretion onto the HeWD through a hot and dense disk.
The ``subduction'' and ``secondary disruption'' pictures are practically identical for our purposes: we follow \citet{Zhang:2013:EarlyRJ} and assume that the HeWD forms the inner layers of the merger remnants' core, with negligible mixing during the merger.
We nevertheless caution that the initial structure of the merger remnant's core is uncertain.}

As in the MS+HeWD case, we first create a HeWD model with mass $M_{\mathrm{HeWD}}$ and $T_{\mathrm{eff}}\approx10\,\mathrm{kK}$ by stripping off a RG model's envelope and allowing the result to cool.
For convergence reasons, we find that we need to synthetically increase the density of the outer layers in order to better match those of the overlying RG core material we wish to add.
To do this, we add a hydrogen-rich envelope in a similar manner to the MS+HeWD.
Once the envelope has reached hydrostatic equilibrium, we then remove it again by excising the envelope layers, producing a HeWD model with artificially dense layers.

We next create a model resembling the layers of the RG following displacement by \textit{a} HeWD \textit{placed at its center}.
To this end, we initialize a star on the helium MS with mass $M_{\mathrm{RGcore}}$ and relax its inner boundary condition to \edit{match} that of the HeWD model.
We then accrete hydrogen-rich material of mass $M_{\mathrm{env}}$ at a high rate $M=10^{-4}M_\odot\,\mathrm{yr}^{-1}$, disabling helium burning and evolving the model until its innermost layer reaches marginal degeneracy, assumed to occur at degeneracy parameter $\eta=1$.
Finally, to create the full merger remnant, we manually stitch together the models representing the \edit{inner} HeWD and the layers of the RG.
As in the MS+HeWD case, we then ring down the merger remnant model for $1\,\mathrm{kyr}$ in \texttt{MESA}'s hydrodynamical mode.
We subsequently disable hydrodynamical mode for the subsequent merger remnant evolution.
The left column of Figure \ref{fig:rg_profiles} shows a RG+HeWD merger remnant constructed using fiducial parameters $M_{\mathrm{HeWD}}=0.40M_\odot$, $M_{\mathrm{RGcore}}=0.20M_\odot$, and $M_{\mathrm{env}}=0.20M_\odot$.

In addition to the fiducial model, we also present a model with $M_{\mathrm{env}}=0.20M_\odot$ and $M_{\mathrm{HeWD}}=0.45M_\odot$ (near maximal-mass HeWD) and a model with $M_{\mathrm{HeWD}}=0.40$ and $M_{\mathrm{env}}=0.40$ (higher-mass envelope).
In both cases, we fix $M_{\mathrm{core}}=0.20M_\odot$, following the finding of \citet{Zhang:2013:EarlyRJ} that core dredge up only occurs for mergers involving high-mass HeWDs and low-mass RG cores.

Despite the considerably simpler procedure of \citet{Zhang:2013:EarlyRJ} involving rapidly accreting material, we find that modern versions of \texttt{MESA} are unable to accommodate it.
Although the cause is uncertain, these convergence issues may have to do with changes to the treatment of accretion in later versions of \texttt{MESA} \citep[see, e.g., Section 7 of][]{Paxton:2015:MESA}.

\section{Results and discussion}

\subsection{Pre-core-dredge-up evolution of MS+HeWD merger remnants}

The evolution of our fiducial MS+HeWD model follows the ``post-dredge-up merger remnant'' outcome of MS+HeWD mergers described in \citetalias{Rui:2024:cv-mergers}.
In summary, the merger remnant begins far up the RGB due to its initially high core mass $M_{\mathrm{HeWD}}=0.40M_\odot$.
After growing its core via hydrogen shell burning, the merger remnant degenerately ignites helium.
Due to its low-entropy core, the shell which first reaches the temperature required for helium burning is much farther off-center than in an isolated RG.
Moreover, by this time, the helium core has grown to a somewhat higher core mass than $\approx0.45M_\odot$ as in an isolated RG \citepalias[see][]{Rui:2024:cv-mergers}.

During its helium flash, the degeneracy of the merger remnant's core is gradually lifted by a series of inward propagating helium subflashes.
As described in \citetalias{Rui:2024:cv-mergers}, these subflashes are atypically energetic and closely spaced in time.
In our fiducial model, the first $\sim10$ subflashes cause the envelope's convective zone to deepen into the helium core, mixing $\approx0.11M_\odot$ of helium-rich material into the envelope.

\subsection{Pre-core-dredge-up evolution of RG+HeWD merger remnants}

\begin{table}
    \centering
    \begin{tabular}{l l l | l l}
        \hline
        $M_{\mathrm{HeWD}}$ & $M_{\mathrm{env}}$ & $M_{\mathrm{core}}$ & $M_{\mathrm{COWD}}$ & $X({}^{22}\mathrm{Ne})_{\mathrm{tot}}$ \\
        \hline
        $0.40$ & $0.20$ & $0.20$ & $0.70$ & $3.27\%$ \\
        $0.45$ & $0.20$ & $0.20$ & $0.73$ & $4.22\%$ \\
        $0.40$ & $0.40$ & $0.20$ & $0.71$ & $2.87\%$ \\
        \hline
    \end{tabular}
    \caption{Initial conditions and final mass and ${}^{22}\mathrm{Ne}$ mass fraction of our RG+HeWD merger remnant models.
    All masses are given in $M_\odot$.}
    \label{tab:rghewd_table}
\end{table}

Initially, the helium core of our fiducial RG+HeWD merger remnant is composed of a hot disrupted RG core overlying a \edit{cool HeWD ($T\simeq10^7\,\mathrm{K}$)}.
Since the core's mass $M_{\mathrm{HeWD}}+M_{\mathrm{RGcore}}$ has contributions from both progenitors, it can substantially exceed the usual core mass $\approx0.45M_\odot$ at which a RG normally experiences the helium flash.

In our fiducial model, the temperature $T\simeq2\times10^8\,\mathrm{K}$ at the base of these disrupted layers arises from the step of our model construction where we allow the RG core to contract in the absence of burning to $\eta=1$ prior to stitching it to the HeWD model.
While the real initial temperature of these layers depends on uncertain details of the merger (e.g., shock heating), we point out that it is plausible for these layers to start off with temperatures greatly exceeding that of helium burning ($T\approx10^8\,\mathrm{K}$).
Although the nuclear heating timescale at the site of helium ignition $t_{\mathrm{heat}}=c_pT/\varepsilon_{\mathrm{nuc}}\sim1\,\mathrm{hr}$ is extraordinarily short, it is still much longer than the core dynamical time ($\sim\mathrm{seconds}$) over which the merger of the RG core and HeWD occurs.
Nuclear burning is therefore not expected to greatly change the temperature profile during the merger.

Due to the high temperatures at the HeWD--RG core interface, degenerate helium burning begins immediately (the dash-dotted curves in the first two columns of Figure \ref{fig:rg_profiles} are identical).
The merger scenario producing our fiducial model involves a relatively high-mass HeWD and a low-mass RG core, conditions which cause helium-flash-driven convection to dredge $\approx0.16M_\odot$ of helium-rich core material into the envelope
\citep{Zhang:2013:EarlyRJ}.
This core dredge-up can be seen by comparing the second and third columns of Figure \ref{fig:rg_profiles}.

\subsection{Post-core-dredge-up evolution and ${}^{22}\mathrm{Ne}$ generation} \label{sect:post-dredge-up}

After experiencing core dredge-up events and starting core helium burning (CHeB), both MS+HeWD and RG+HeWD merger remnants subsequently evolve similarly.
In addition to enriching the envelope in helium, the core dredge-up also produces enhancements in other species such as ${}^{12}\mathrm{C}$, transforming the star into a carbon star ($\mathrm{C}/\mathrm{O}>1$).
On the early AGB, as the hydrogen shell burning front grows the core and moves to larger mass coordinates, this ${}^{12}\mathrm{C}$ is largely converted to ${}^{14}\mathrm{N}$ in the presence of CNO-cycle hydrogen burning.
In this form, ${}^{14}\mathrm{N}$ is quickly processed into ${}^{22}\mathrm{Ne}$ via the following reaction chain involving two $\alpha$ captures and a $\beta$ decay:
\begin{subequations}
    \begin{gather}
        {}^{14}\mathrm{N} + {}^{4}\mathrm{He} \rightarrow {}^{18}\mathrm{F} + \gamma \\
        {}^{18}\mathrm{F} \rightarrow {}^{18}\mathrm{O} + e^+ + \nu_e \\
        {}^{18}\mathrm{O} + {}^{4}\mathrm{He} \rightarrow {}^{22}\mathrm{Ne} + \gamma\mathrm{.}
    \end{gather}
\end{subequations}

\begin{figure}
    \centering
    \includegraphics[width=\linewidth]{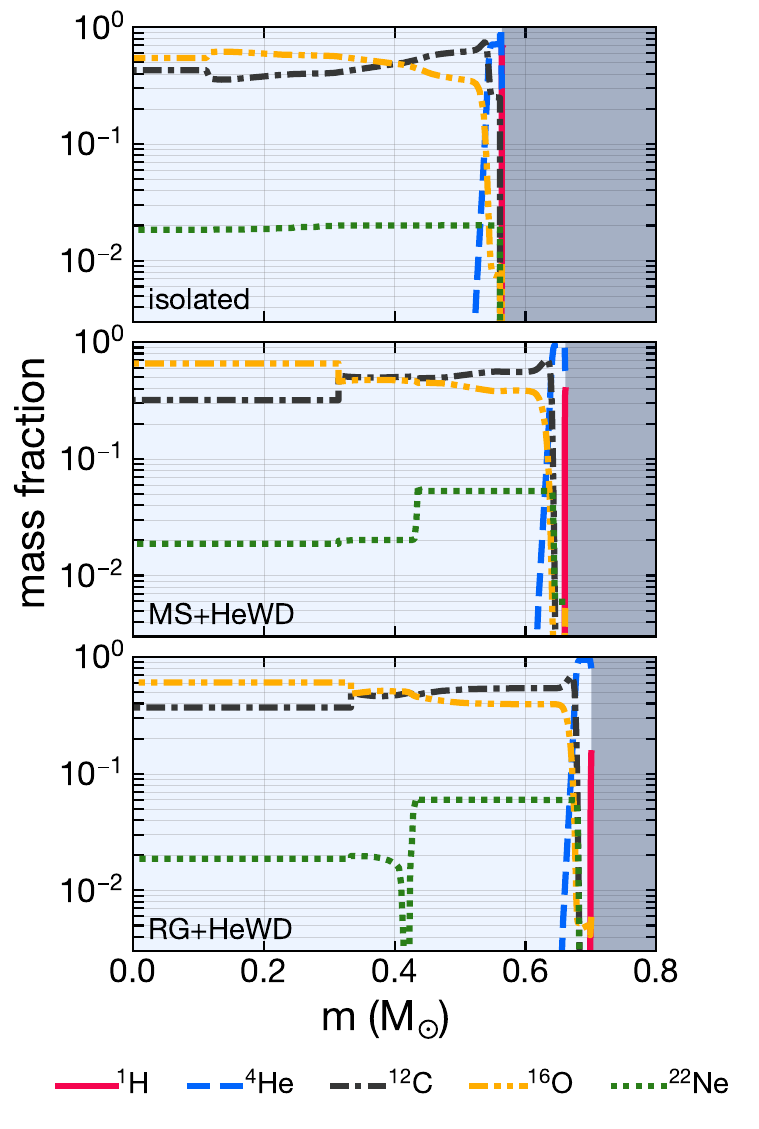}
    \caption{Composition profiles of an isolated RG, fiducial MS+HeWD merger remnant (Section \ref{sect:making_mshewd}), and fiducial RG+HeWD merger remnant (Section \ref{sect:making_rghewd}) after expulsion of their hydrogen-rich envelopes at the beginning of their COWD cooling tracks ($\log g=6.0$).}
    \label{fig:comp_early_cowd}
\end{figure}

As a result, once the merger remnants have expelled their envelopes following the AGB, their outer layers are enriched in ${}^{22}\mathrm{Ne}$ (Figure \ref{fig:comp_early_cowd}).
The total ${}^{22}\mathrm{Ne}$ mass fractions in our fiducial MS+HeWD and RG+HeWD merger remnant models reach $\approx3.0\%$ and $\approx3.3\%$, respectively (cf. $\approx1.9\%$ for our single COWD model).
If $X({}^{22}\mathrm{Ne})\simeq3\%$ is adopted as the criterion required for ${}^{22}\mathrm{Ne}$ distillation, the fiducial RG+HeWD merger remnant comfortably distills ${}^{22}\mathrm{Ne}$, whereas the fiducial MS+HeWD merger remnant is right on the edge between standard CO crystallization and ${}^{22}\mathrm{Ne}$ distillation.
However, the precise threshold required for ${}^{22}\mathrm{Ne}$ distillation depends on the carbon-to-oxygen ratio \citep{Blouin:2021:QBranch}, which is mainly set by the notoriously uncertain ${}^{12}\mathrm{C}(\alpha,\gamma)^{16}\mathrm{O}$ reaction rate \citep[e.g.,][]{Holt:2019:COReactionRate}.

Due to their increased mean molecular weights, the merger remnants' now helium-enriched envelopes burn hydrogen more vigorously \citep{Refsdal:1970:CoreMassLuminosityRelation}.
As discussed in \citetalias{Rui:2024:cv-mergers}, this causes the remnants' helium cores to grow by several tenths of a solar mass during CHeB, rather than only a few hundredths as in the isolated RG model (compare the third and fourth columns of Figure \ref{fig:rg_profiles}).
In particular, the cores of the fiducial MS+HeWD and RG+HeWD models grow by $\approx0.22M_\odot$ and $\approx0.25M_\odot$ during CHeB, respectively.
Even though the core dredge-up reduces the initial core mass, both merger remnant models' cores end up with higher masses than typical CHeB stars.
These slightly overmassive cores are inherited by the resulting COWDs.
In particular, our fiducial MS+HeWD and RG+HeWD models produce final COWD masses of $\approx0.66M_\odot$ and $\approx0.70M_\odot$, compared to $\approx0.56M_\odot$ in our isolated model.

\begin{figure}
    \centering
    \includegraphics[width=\linewidth]{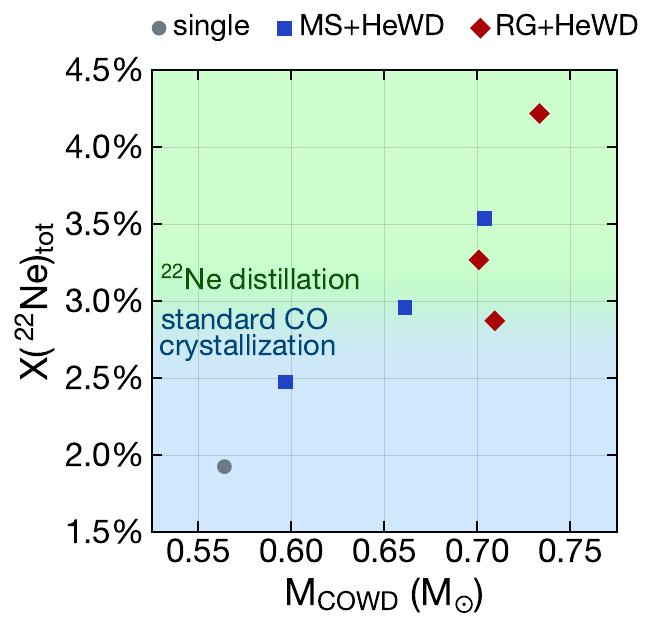}
    \caption{Final masses $M_{\mathrm{COWD}}$ and total ${}^{22}\mathrm{Ne}$ mass fractions $X({}^{22}\mathrm{Ne})_{\mathrm{tot}}$ of the COWDs formed by our merger remnant models.}
    \label{fig:xne_summary}
\end{figure}

The precise masses and amount of ${}^{22}\mathrm{Ne}$ generated depends somewhat on the initial conditions of the stellar merger (Figure \ref{fig:xne_summary}).
Compared to our fiducial MS+HeWD case, our MS+HeWD merger remnant model with an increased $M_{\mathrm{HeWD}}=0.45M_\odot$ creates a more massive COWD ($M_{\mathrm{COWD}}\approx0.70M_\odot$) with a larger $X({}^{22}\mathrm{Ne})\approx3.5\%$ (Table \ref{tab:mshewd_table}).
This is because the larger core-dredge-up event caused by the more massive HeWD produces a greater enhancement in ${}^4\mathrm{He}$ and ${}^{12}\mathrm{C}$.
The former of these increases the growth of the core whereas the latter is the main ingredient for the production of ${}^{22}\mathrm{Ne}$.
Conversely, our MS+HeWD model with an increased $M_{\mathrm{env}}=0.60M_\odot$ results in a less massive COWD ($M_{\mathrm{COWD}}\approx0.60M_\odot$) and a lower $X({}^{22}\mathrm{Ne})\approx2.5\%$.
The more massive envelope more effectively dilutes the impact of the core dredge up.
The dilution effect is usually more important than the overall larger mass of the remnant in setting the final mass of the COWD.

The impact of increasing $M_{\mathrm{HeWD}}=0.45M_\odot$ is also to increase both $M_{\mathrm{COWD}}\approx0.73M_\odot$ and $X({}^{22}\mathrm{Ne})\approx4.2\%$ (Table \ref{tab:rghewd_table}).
As in the MS+HeWD merger remnant case, this is because of the increased mass dredged up from the core.
However, when increasing the envelope mass to $M_{\mathrm{env}}=0.40M_\odot$, the final COWD mass increases slightly to $M_{\mathrm{COWD}}\approx0.71M_\odot$, although $X({}^{22}\mathrm{Ne})\approx2.9\%$ decreases as in the MS+HeWD case.
The increased COWD mass is due to the prolonged thermally pulsing AGB phase during which the CO core can continue to grow.
In this case, this effect marginally wins out over the diluting effect of the more massive envelope.
The close relationship between dredged-up core mass and both $M_{\mathrm{COWD}}$ and $X({}^{22}\mathrm{Ne})$ is responsible for the positive correlation between $M_{\mathrm{COWD}}$ and $X({}^{22}\mathrm{Ne})$ in Figure \ref{fig:xne_summary}.
These models demonstrate that, under some (but not all) conditions, merger remnant COWDs can be ${}^{22}\mathrm{Ne}$-rich \edit{enough} to distill ${}^{22}\mathrm{Ne}$.

\subsection{The role of thermohaline mixing} \label{sect:thermohaline}

Although our merger remnants can generate enough ${}^{22}\mathrm{Ne}$ to enable ${}^{22}\mathrm{Ne}$ distillation ($X({}^{22}\mathrm{Ne})\gtrsim3\%$), most of this ${}^{22}\mathrm{Ne}$ is concentrated in the merger remnants' outer layers at the start of the COWD phase (Figure \ref{fig:comp_early_cowd}).
\edit{${}^{22}\mathrm{Ne}$ distillation begins immediately upon crystallization only if enough of this ${}^{22}\mathrm{Ne}$ ($X({}^{22}\mathrm{Ne})\gtrsim3\%$) can be mixed into the center.}
Otherwise, standard CO crystallization occurs until the crystallization front reaches a layer with $X({}^{22}\mathrm{Ne})\gtrsim3\%$, and ${}^{22}\mathrm{Ne}$ distillation occurs for a shorter time $\simeq2\,\mathrm{Gyr}$ \citep{Blouin:2021:QBranch}.
The latter scenario is qualitatively similar to what is expected to occur in single COWDs.

\begin{figure}
    \centering
    \includegraphics[width=\linewidth]{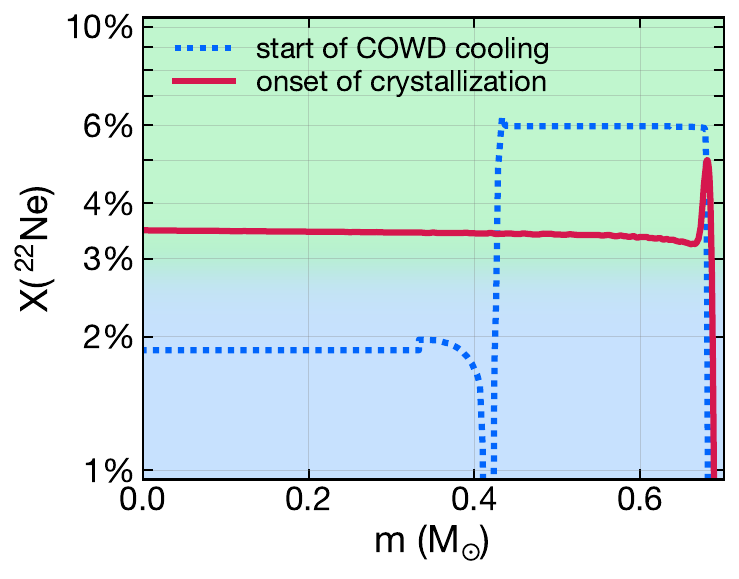}
    \caption{The mass fraction profile of ${}^{22}\mathrm{Ne}$ in the COWD resulting from the fiducial RG+HeWD merger remnant.
    The profile is shown both for the beginning of the COWD phase and \edit{at the} onset of CO crystallization\edit{, by which time thermohaline mixing has fully homogenized its internal composition}.}
    \label{fig:thermohaline_ne22}
\end{figure}

In our models, downward mixing of ${}^{22}\mathrm{Ne}$ into the core of the COWD is dominated by thermohaline mixing.
Thermohaline mixing is a type of double-diffusive convection in which stable thermal (and overall) stratification is destabilized by an unstable composition gradient \citep[e.g.,][]{Charbonnel:2007:ThermohalineRGB}, here supplied by the higher concentration of ${}^{22}\mathrm{Ne}$ in the outer layers.
\edit{We evolve our fiducial RG+HeWD model through the COWD cooling track including thermohaline mixing through the classic prescription of \citet{Kippenhahn:1980:Thermohaline}, setting $\alpha_{\mathrm{th}}=2.0$ \citep[following][]{Cantiello:2010:Thermohaline} and evaluating convective stability using the Ledoux criterion \citep{Ledoux:1947:LedouxCriterion}.
We also include the effects of gravitational settling, using eight species as diffusion representatives: ${}^{1}\mathrm{H}$, ${}^{3}\mathrm{He}$, ${}^{4}\mathrm{He}$, ${}^{12}\mathrm{C}$, ${}^{16}\mathrm{O}$, ${}^{20}\mathrm{Ne}$, ${}^{22}\mathrm{Ne}$, and ${}^{26}\mathrm{Mg}$.
For simplicity, we use the default \texttt{T\_tau} atmosphere option in MESA, and do not include any convective overshoot.
Figure \ref{fig:thermohaline_ne22} shows the COWD composition profile both at the beginning of the cooling sequence and at the onset of crystallization.
}

\edit{We find that thermohaline mixing is comfortably efficient enough to fully homogenize the COWD's interior by the time it has cooled to $T_{\mathrm{eff}}\simeq20\,\mathrm{kK}$, implying that prompt ${}^{22}\mathrm{Ne}$ distillation will occur.
Once the COWD cools below $\log(T_{\mathrm{eff}}/\mathrm{K})=4.2$ ($T_{\mathrm{eff}}\lesssim15.8\,\mathrm{kK}$), we disable the Ledoux criterion to avoid problematic and unphysical admixture of helium into the degenerate interior.
The COWD model then begins crystallization once it has cooled to $T_{\mathrm{eff}}\approx7.3\,\mathrm{kK}$, consistent with the widely used STELUM models \citep{Bedard:2020:STELUM} which predict a crystallization temperature $T_{\mathrm{eff}}\approx7.0\,\mathrm{kK}$ ($T_{\mathrm{eff}}\approx7.7\,\mathrm{kK}$) for a $0.70M_\odot$ COWD with a thick (thin) hydrogen atmosphere.
}

\edit{While the efficiency of thermohaline mixing is uncertain, our implementation represents a conservative estimate of its strength.
Standard thermohaline mixing prescriptions, while comparable in strength to newer prescriptions calibrated to three-dimensional simulations \citep{Traxler:2011:Thermohaline,Brown:2013:Thermohaline}, are roughly two orders of magnitude too weak to explain observed RG abundances \citep[][see also \citealt{Ulrich:1972:Thermohaline}]{Charbonnel:2007:ThermohalineRGB,Cantiello:2010:Thermohaline}.
}
More recent simulations have revealed that even small vertical magnetic fields $\sim300\,\mathrm{G}$ can dramatically enhance the efficiency of thermohaline mixing up to the point of compatibility of RG observations \citep{Harrington:2019:MagneticThermohaline}.
The basic mechanism is that vertical magnetic fields \edit{provide magnetic stress which reinforces the growing instability against parasitic instabilities \citep{Holyer:1984:Thermohaline,Radko:2012:ParasiticInstabilities,Brown:2013:Thermohaline}, allowing the flow to achieve a saturated state} which transport\edit{s} heat and chemicals with much higher efficiency \citep[although this picture is complicated by non-ideal effects;][]{Fraser:2024:MagneticThermohaline}.
\edit{Our merger remnants, which were plausibly magnetized by the mergers which produced them \citep[e.g.,][]{Schneider:2016:MagneticMergers,Ohlmann:2016:CEEMagnetic}, have thermohaline mixing plausibly in this regime and may homogenize even faster than in our models.}

\subsection{Very late thermal pulses and the spectral type of the COWD} \label{sect:vltps}

During the AGB, hydrogen shell burning continually deposits helium ashes on the carbon--oxygen core.
This thin helium layer unstably burns in periodic thermal pulses whenever it accumulates to sufficiently high masses.
Very late thermal pulses (which occur once most hydrogen burning has ceased) can mix residual hydrogen into the helium-burning layer \citep{Iben:1983:LateThermalPulse}.
This is thought to burn up any remaining hydrogen and leave the COWD with a helium atmosphere (spectral types DO and DB).
In typical stars, they are expected to happen a minority of the time.
Our representative single star model ends its life as a COWD with near-canonical hydrogen and helium masses $M_{\mathrm{H}}\approx6\times10^{-5}M_\odot$ and $M_{\mathrm{He}}\approx1.7\times10^{-2}M_\odot$.
Following the typical case, this model does not experience a late thermal pulse and subsequently cools as a hydrogen-atmosphere (DA) COWD.

Curiously, most of our post-core-dredge-up MS+HeWD and RG+HeWD merger remnants experience a very late thermal pulse.
Roughly $\simeq10^5\,\mathrm{yr}$ after the end of the AGB phase, these very late thermal pulses quickly burn the initially high-mass helium layers ($M_{\mathrm{He}}>2\times10^{-2}M_\odot$), depleting them to $M_{\mathrm{He}}\lesssim10^{-2}M_\odot$.
We suspect the frequent occurrence of late thermal pulses is driven by the more rapid hydrogen burning caused by the increased mean molecular weight, as discussed in Section \ref{sect:post-dredge-up}.

Whether these very late thermal pulses can fully destroy the hydrogen depends on the strength of convective overshoot.
This, in turn, sets whether the spectral type of the COWD is DA (hydrogen-atmosphere) or DB (helium-atmosphere) by the time of crystallization.
We test the impact of convective overshoot by evolving our fiducial RG+HeWD merger remnant through the very late thermal pulse, both with and without convective overshoot.
Convective overshoot is modeled as exponential overmixing with scale height $f_{\mathrm{ov}}H_p=0.015H_p$, with the overshooting region starting $0.005H_p$ inside the convective zone.

Our fiducial COWD starts with a near-canonical hydrogen mass $M_{\mathrm{H}}\approx1.3\times10^{-4}M_\odot$.
If no convective overshoot is included, \edit{$M_{\mathrm{H}}\approx3.3\times10^{-6}M_\odot$} survives the very late thermal pulse \citep[cf.][]{Iben:1983:LateThermalPulse}.
Due to gravitational settling, this hydrogen accumulates in the COWD's surface layers, which become essentially pure hydrogen.
Although most hydrogen is destroyed, the remaining hydrogen is more than enough to prevent mixing into the COWD's helium layer when the near-surface convective zone deepens to \edit{$M_{\mathrm{conv}}\lesssim10^{-8}M_\odot$} by the time of crystallization \citep[$T_{\mathrm{eff}}\simeq7.5\,\mathrm{kK}$;][]{Tremblay:2015:DAWDConvection}.
In this case, the COWD is of spectral type DA by the time of crystallization, and our merger scenarios are a promising channel for producing DAHe-type COWDs (see Section \ref{sect:candidates}).

In contrast, when convective overshoot is included, hydrogen more efficiently mixes into the helium burning layer during the very late thermal pulse, where it is destroyed.
The COWD is left with a tiny hydrogen mass \edit{$M_{\mathrm{H}}\approx4.9\times10^{-11}M_\odot$}, with even the outermost layers dominated by helium (\edit{$X\approx3.1\times10^{-5}$}) by the time of crystallization \edit{(which now occurs at a slightly higher $T_{\mathrm{eff}}\approx9.1\,\mathrm{kK}$)}.
This dependence of the surviving hydrogen mass on the convective overshoot is consistent with the results of \citet{Herwig:1999:VLTPOvershoot} and \citet{Herwig:2000:VLTPOvershoot}.
\edit{Although COWDs with $M_{\mathrm{H}}\gtrsim10^{-12}M_\odot$ can still retain hydrogen atmospheres \citep[e.g.,][]{Tremblay:2015:DAWDConvection}, in our model the hydrogen is well-mixed within the helium-partial-ionization-driven convective zone \citep[cf.][who suggest that all DBs may have trace amounts of hydrogen]{Koester:2015:DBAs}.}
If convective overshoot is \edit{sufficient to change} the spectral type of the COWD to DB, our models would predict the existence of somewhat overmassive DB-type COWDs which undergo ${}^{22}\mathrm{Ne}$-distillation-related phenomena (e.g., DBHe-type COWDs).
However, \edit{\citet{Ginzburg:2025:DoubleFacedWDs}} find that magnetic fields $\gtrsim1\,\mathrm{MG}$ may inhibit the development of near-surface convection zones.
If merger remnant COWDs are born sufficiently magnetic \citep[e.g., as the result of merger;][]{Schneider:2016:MagneticMergers,Ohlmann:2016:CEEMagnetic}, magnetic fields may inhibit convection enough to allow hydrogen to gravitationally settle to the surface rather than being convectively mixed into the helium layer.
This may allow them to retain their hydrogen atmospheres and retain their DA spectral types.

\subsection{Comparison to candidate ${}^{22}\mathrm{Ne}$-distilling WDs} \label{sect:candidates}

The Q branch is a cooling-delay-induced overdensity of ultramassive ($\gtrsim1.1M_\odot$) WDs in the Hertzsprung--Russell diagram \citep{GaiaCollaboration:2018:HRDQBranch}.
Although isolated ultramassive WDs are expected to have oxygen--neon cores \citep{Nomoto:1984:ONeWDs,GarciaBerro:1994:SuperAGB}, the Q branch follows the predicted location of CO crystallization \citep[e.g.,][]{Bedard:2020:STELUM}.
Curiously, \citet{Cheng:2019:QBranch} show that the Q branch population requires that $\simeq6\%$ of WDs in the Q branch must experience a very long cooling delay $\simeq8\,\mathrm{Gyr}$ which cannot be supplied by standard CO crystallization.
Q branch WDs are thus prime candidates for WDs undergoing ${}^{22}$Ne distillation \citep{Blouin:2021:QBranch,Bedard:2024:Buoyant}.

Although MS+HeWD and RG+HeWD merger remnants can generate enough ${}^{22}\mathrm{Ne}$ to allow ${}^{22}\mathrm{Ne}$ distillation to occur, they are unlikely to solve the Q branch anomaly.
While COWDs from these channels can be somewhat more massive than typical COWDs, they do not approach the ultramassive regime.
In both cases, $M_{\mathrm{WD}}$ is limited to $\simeq0.45M_\odot$ (the maximum mass of a HeWD).
While larger values of $M_{\mathrm{env}}$ (i.e., more massive MS or RG progenitors) increase the total mass available to grow the COWD, they also dilute the dredged-up helium responsible for increasing the core's growth rate as well as the dredged-up carbon which ultimately forms the ${}^{22}\mathrm{Ne}$ in the first place.
While MS/RG+HeWD mergers cannot plausibly form ultramassive Q branch WDs, they are capable of forming lower-mass ``cousins'' to this population.

Another anomalous class of observed hydrogen-atmosphere WDs (of type DAHe/DAe) display anomalous Balmer emission lines.
At present, approximately two dozen have been discovered \citep{Greenstein:1985:GD356,Reding:2020:DAHe,Gansicke:2020:DAHe,Walters:2021:DAHe,Farihi:2023:DAHe,Manser:2023:DESIDAHe}.
Strikingly, besides sharing emission-line phenomenology, DA(H)e WDs are mostly confirmed to be magnetic and, like Q branch WDs, closely cluster around the CO crystallization line (at $T_{\mathrm{eff}}\simeq7500\,\mathrm{K}$).
They also tend to be rapidly rotating and somewhat more massive than typical WDs ($\simeq0.8M_\odot$).
The shared characteristics of these WDs and the lack of detected close companions \citep{Ferrario:1997:DAHe,Wickramasinghe:2010:DAHe} strongly suggest the emission lines arise from an as-of-yet mysterious source of atmospheric heating.
In this context, \citet{Lanza:2024:DAHes} propose that the required atmospheric heating can be caused by the resistive dissipation of currents powered by a ${}^{22}\mathrm{Ne}$ distillation-driven dynamo.
DAHe WDs' somewhat high masses and fast rotation rates suggest that they are the remnants of mergers.
Such mergers may be responsible for the high required ${}^{22}\mathrm{Ne}$ abundances required for a ${}^{22}\mathrm{Ne}$-distillation-driven dynamo to operate.

MS/RG+HeWD mergers are a plausible formation channel for DAHe WDs.
Our merger remnant models produce COWDs which have masses $\simeq0.1$--$0.2\,M_\odot$ greater than our isolated COWD model (with mass $\approx0.56M_\odot$).
\edit{DAHe WDs cluster around masses $\simeq0.8M_\odot$ with a spread $\simeq0.1M_\odot$ \citep[see Figure 7 of][]{Manser:2023:DESIDAHe}.
Although none of our models formally reach as high as $\simeq0.8M_\odot$, the true final COWD mass} may depend somewhat on stellar-evolution uncertainties such as mixing \citep[e.g.,][]{Constantino:2017:CHeBMixing} and may be increased by additional admixture of helium into the merger remnant RG's envelope (e.g., during the merger itself\edit{, or due to convective overshoot during the core dredge-up event}).
As merger remnants, it is also physically plausible for them to be rapidly rotating, although this depends on whether angular momentum transport is efficient enough to slow the core of the remnant before it becomes a COWD \citep[e.g.,][]{Fuller:2019:TS-AMT}.
This scenario also requires that the hydrogen atmosphere at least sometimes survives the very late thermal pulse, which depends on the strength of convective overshoot (see Section \ref{sect:vltps}).

\citet{Izzard:2007:EarlyR} and \citet{Zhang:2013:EarlyRJ} show that RG+HeWD mergers are capable of forming carbon-rich core-helium-burning stars, in particular early R-type carbon stars.
MS+HeWD merger remnants, which undergo similar core dredge ups \citepalias{Rui:2024:cv-mergers}, are also plausible progenitors.
If both DAHe WDs and early-R \edit{stars} really descend from MS/RG+HeWD merger remnants which have undergone core-dredge up-events, their volumetric birth rates should be consistent.
If the current sample of $\approx24$ known DAHe WDs is roughly complete to $\simeq150\,\mathrm{pc}$ \citep[see the discussion in the Conclusion Section of][]{Lanza:2024:DAHes}, their space density is $n_{\mathrm{DAHe}}\sim2\times10^{-6}\,\mathrm{pc}^{-3}$.
If cooling in DAHe WDs is stalled for $\tau_{\mathrm{DAHe}}\sim8\,\mathrm{Gyr}$, this implies a DAHe volumetric birth rate $n_{\mathrm{DAHe}}/\tau_{\mathrm{DAHe}}\sim2\times10^{-7}\,\mathrm{pc}^{-3}\,\mathrm{Gyr}^{-1}$.
On the other hand, for a typical core-helium-burning lifetime $T_{\mathrm{earlyR}}\sim100\,\mathrm{Myr}$ and a space density of early R stars $n_{\mathrm{earlyR}}\sim(1.7$--$15)\times10^{-8}\,\mathrm{pc}^{-3}$ given by \citet{Knapp:2001:EarlyR} and \citet{Bergeat:2002:EarlyR}, the early-R-star volumetric birth rate is $n_{\mathrm{earlyR}}/\tau_{\mathrm{earlyR}}\sim(1.7$--$15)\times10^{-7}\,\mathrm{pc}^{-3}\,\mathrm{Gyr}^{-1}$.
The agreement between these two rates promisingly suggests that DAHes may descend in large part from MS/RG+HeWD mergers.

\subsection{Comparison to the COWD+subgiant merger scenario}

\citet{Shen:2023:Subgiant} propose that Q branch COWDs result from the merger of a massive ($\simeq1M_\odot$) COWD and a subgiant.
In this scenario, dredged-up carbon from the COWD burns in the hot outer layers of the merger remnant.
In these hydrogen- and helium-rich conditions, this burning eventually enhance the outer layers of the final COWD in ${}^{22}\mathrm{Ne}$ and ${}^{26}\mathrm{Mg}$.
\edit{Thermohaline mixing then transports both of these neutron-rich species to the center of the final COWD.}
Unlike in our scenarios, the COWD+subgiant scenario can plausibly explain ${}^{22}\mathrm{Ne}$-distillation-related phenomena for ultramassive WDs due to its flexibility in the mass of the original COWD.
For lower-mass initial COWDs $\simeq0.6M_\odot$, COWD+subgiant mergers may also be capable of producing ${}^{22}\mathrm{Ne}$-distilling COWDs in the mass range of DAHe COWDs.
In their scenario, the final COWD undergoes distillation immediately upon crystallization due to the increases in mean molecular weight of the liquid phase by both ${}^{22}\mathrm{Ne}$ and ${}^{26}\mathrm{Mg}$.
In contrast, in both our MS+HeWD and RG+HeWD scenarios, nuclear burning does not occur at high enough temperatures to produce significant amounts of ${}^{26}\mathrm{Mg}$.
Instead, our scenarios simply produce a greater amount of ${}^{22}\mathrm{Ne}$ overall, sufficient to enable distillation.

\section{Conclusions}

This work demonstrates two formation channels for COWDs with ${}^{22}\mathrm{Ne}$ abundances high enough to undergo ${}^{22}\mathrm{Ne}$ distillation.
Both channels involve stellar engulfments of HeWDs, and produce the required ${}^{22}\mathrm{Ne}$ enhancement ($X({}^{22}\mathrm{Ne})\gtrsim3\%$) through nuclear processing of ${}^{12}\mathrm{C}$ dredged up following abnormally energetic and off-center helium ignition events.
Our chief findings are as follows:
\begin{itemize}
    \item The remnants of MS/RG+HeWD mergers can evolve into ${}^{22}\mathrm{Ne}$-distilling COWDs with somewhat higher total masses (in our models, up to $\approx0.73M_\odot$).
    
    \item The final mass $M_{\mathrm{COWD}}$ and ${}^{22}\mathrm{Ne}$ abundance $X({}^{22}\mathrm{Ne})_{\mathrm{tot}}$ are both positively correlated with the amount of core material dredged up during helium ignition.
    This causes more ${}^{22}\mathrm{Ne}$-rich merger remnant COWDs to also have higher masses.
\end{itemize}

Our predicted COWD masses and merger rates both suggest MS/RG+HeWD mergers as a promising formation channel for DAHe-type COWDs, although these channels cannot produce COWDs massive enough to explain the ultramassive Q-branch cooling anomaly.
This hypothesis comes with the following caveats:

\begin{itemize}

    \item \edit{Our merger remnant models only undergo the required RG-like evolution if we construct their initial profiles assuming the retention of a significant amount of hydrogen during the merger.
    Following the merger itself in greater detail may be necessary to fully assess how much hydrogen can be successfully incorporated into the merger remnant.
    In particular, in the MS+HeWD case, simulations of the highly super-Eddington disk formed from the disrupted MS component suggest that most hydrogen should be expelled from the system \citep{Metzger:2021:cv-deaths}.}

    \item \edit{Our merger remnant COWDs can be somewhat higher-mass than typical COWDs ($\simeq0.6M_\odot$) due to enhanced hydrogen burning in helium-rich shells.
    However, our models do not reproduce the full range of observed DAHe WD masses $\simeq0.6$--$0.9M_\odot$ \citep{Manser:2023:DESIDAHe}, with our most massive COWD model reaching $\approx0.73M_\odot$.
    This tension may be resolved by additional mixing of helium into the hydrogen envelope, such as during the merger itself or due to convective overshoot during helium ignition.
    Since our models ignore these effects, they likely underestimate the final COWDs masses that can be produced.}

    \item Most MS/RG+HeWD merger remnants undergo a very late thermal pulse during the COWD phase.
    Depending on the strength of convective overshoot and the subsequent behavior of near-surface convection zones, the hydrogen atmosphere may be destroyed, leaving a COWD of spectral type DB.
    However, inhibition of near-surface convection by magnetic fields $\gtrsim1\,\mathrm{MG}$ may be sufficient for the retention of the hydrogen atmosphere even if convective overshoot is efficient \citep{Ginzburg:2025:DoubleFacedWDs}.
\end{itemize}

While we demonstrate that MS/RG+HeWD mergers are a plausible channel for some ${}^{22}\mathrm{Ne}$-distilling WDs, many uncertainties remain.
Although this work models the stellar evolution of the merger remnants in detail, the structure of the initial merger remnant depends on the detailed hydrodynamics of the merger itself.
Similarly, the spin of the final COWD depends sensitively on the physics of angular momentum transport, which is also not particularly well understood, even in the case of single stars \citep{Aerts:2019:AMT}.
Finally, while we have presented a handful of exploratory stellar models, a more detailed parameter study is required to fully assess the diversity of COWD outcomes and make more precise rate estimates.

{\,}\\

\noindent We thank \edit{Evan Bauer,} Ebraheem Farag, Jared Goldberg, Philip Mocz, Rich Townsend, and Sunny Wong for technical support for the \texttt{MESA} code, and Sivan Ginzburg, Nuccio Lanza, Jay Farihi, \edit{and the anonymous referee} for helpful scientific comments.
N.Z.R. acknowledges support from the National Science Foundation Graduate Research Fellowship under Grant No. DGE‐1745301, from the United States--Israel Binational Science Foundation through grant BSF-2022175, and through the NASA Hubble Fellowship grant HST-HF2-51589.001-A awarded by the Space Telescope Science Institute, which is operated by the Association of Universities for Research in Astronomy, Inc., for NASA, under contract NAS5-26555. Finally, we \edit{dedicate this work to} Bill Paxton\edit{,} for his enormous contributions to the field of stellar physics through his development of the \texttt{MESA} code, democratizing the study of stars and enabling this work and countless others.

\facilities{None}

\software{\texttt{MESA} \citep{Paxton:2011:MESA,Paxton:2013:MESA,Paxton:2015:MESA,Paxton:2018:MESA,Paxton:2019:MESA,Jermyn:2023:MESA}}

\bibliography{bibliography}{}
\bibliographystyle{aasjournalv7}

\end{document}